\documentstyle[11pt,adassconf]{article}  
\begin{document}   
\paperID{O1-02}

\title{The Sloan Digital Sky Survey$^1$ and its Archive}

\author{Alexander S. Szalay, Peter Kunszt, Anirudha Thakar}
\affil{Department of Physics and Astronomy, 
	The Johns Hopkins University, Baltimore, MD 21218} 
\author{Jim Gray and Don Slutz}
\affil{Microsoft Research, San Francisco, CA 94105}

\altaffiltext{1}{The Sloan Digital Sky Survey (SDSS) is a joint
project of The University of Chicago, Fermilab, the Institute for
Advanced Study, the Japan Participation Group, The Johns Hopkins
University, the Max-Planck-Institute for Astronomy, Princeton
University, the United States Naval Observatory, and the University of
Washington. Apache Point Observatory, site of the SDSS, is operated by
the Astrophysical Research Consortium. Funding for the project has
been provided by the Alfred P. Sloan Foundation, the SDSS member
institutions, the National Aeronautics and Space Administration, the
National Science Foundation, the U.S. Department of Energy and the
Ministry of Education of Japan. The SDSS Web site is
http://www.sdss.org/ }

\contact{Alex Szalay}
\email{szalay@jhu.edu}

\paindex{Szalay, A. S.}
\aindex{Kunszt, P.}     
\aindex{Thakar, A.}
\aindex{Gray, J.}
\aindex{Slutz, D.}     


\begin{abstract}

The next-generation astronomy archives will cover most of the universe
at fine resolution in many wavelengths. One of the first of these
projects, the Sloan Digital Sky Survey (SDSS) will create a
5-wavelength catalog over 10,000 square degrees of the sky. The 200
million objects in the multi-terabyte database will have mostly
numerical attributes, defining a space of 100+ dimensions.  Points in
this space have highly correlated distributions. The archive will
enable astronomers to explore the data interactively. Data access will
be aided by multidimensional spatial indices. The data will be
partitioned in many ways. Small tag objects consisting of the most
popular attributes speed up frequent searches.  Splitting the data
among multiple servers enables parallel, scalable I/O.  Hashing
techniques allow efficient clustering and pairwise comparison
algorithms. Randomly sampled subsets allow debugging otherwise large
queries at the desktop. Central servers will operate a data pump that
supports sweeping searches that touch most of the data. 

\end{abstract}

\section{Introduction}

Astronomy is undergoing a major paradigm shift.  Data gathering
technology is riding Moore's law: data volumes are doubling quickly,
and becoming more homogeneous. For the first time data acquisition and
archival is being designed for online interactive analysis.  Shortly,
it will be much easier to download a detailed sky map or object class
catalog, than wait several months to access a telescope that is often
quite small.  Several multi-wavelength projects are under way: SDSS,
GALEX, 2MASS, GSC-2, POSS2, ROSAT, FIRST and DENIS, each surveying
a large fraction of the sky. Together they will yield a Digital Sky,
of interoperating multi-terabyte databases.  In time, more catalogs
will be added and linked to the existing ones.  Query engines will
become more sophisticated, providing a uniform interface to all these
datasets.  In this era, astronomers will have to be just as familiar
with mining data as with observing on telescopes.

\section{The Sloan Digital Sky Survey}

The Sloan Digital Sky Survey (SDSS) will digitally map about half of
the Northern sky in five spectral bands from ultraviolet to the near
infrared. It is expected to detect over 200 million
objects. Simultaneously, it will measure redshifts for the brightest
million galaxies (see http://www.sdss.org/).  The SDSS is the
successor to the Palomar Observatory Sky Survey (POSS), which provided
a standard reference data set to all of astronomy for the last 40
years. Subsequent archives will augment the SDSS and will interoperate
with it. The SDSS project thus consists of not only of building the
hardware, and reducing and calibrating the data, but also includes
software to classify, index, and archive the data so that many
scientists can use it.  The SDSS will revolutionize astronomy,
increasing the amount of information available to researchers by
several orders of magnitude.  The SDSS archive will be large and
complex: including textual information, derived parameters, multi-band
images, spectra, and temporal data. The catalog will allow astronomers
to study the evolution of the universe in great detail.  It is
intended to serve as the standard reference for the next several
decades.  After only a month of operation, SDSS found the two most
distant known quasars.  With more data, other exotic properties will
be easy to mine from the datasets. The potential scientific impact of
the survey is stunning.  To realize this potential, data must be
turned into knowledge. This is not easy - the information content of
the survey will be larger than the entire text contained in the
Library of Congress.

The SDSS is a collaboration between the University of Chicago,
Princeton University, the Johns Hopkins University, the University of
Washington, Fermi National Accelerator Laboratory, the Japanese
Participation Group, the United States Naval Observatory, and the
Institute for Advanced Study, Princeton, with additional funding
provided by the Alfred P. Sloan Foundation, NSF and NASA. The SDSS
project is a collaboration between scientists working in diverse areas
of astronomy, physics and computer science. 

The survey will be carried out with a suite of tools developed and
built especially for this project - telescopes, cameras, fiber
spectrographic systems, and computer software.  SDSS constructed a
dedicated 2.5-meter telescope at Apache Point, New Mexico, USA. The
telescope has a large, flat focal plane that provides a 3-degree field
of view. This design balances the areal coverage of the instrument
against the detector's pixel resolution.  The survey has two main
components: a photometric survey, and a spectroscopic survey. The
photometric survey is produced by drift scan imaging of 10,000 square
degrees centered on the North Galactic Cap using five broadband
filters that range from the ultraviolet to the infrared. The effective
exposure is 55 sec. The photometric imaging uses an array of 30x2Kx2K
Imaging CCDs, 22 Astrometric CCDs, and 2 Focus CCDs. Its 0.4 arcsec
pixel size provides a full sampling of the sky. The data rate from the
120 million pixels of this camera is 8 Megabytes per second. The
cameras can only be used under ideal conditions, but during the 5
years of the survey SDSS will collect more than 40 Terabytes of image
data.  The spectroscopic survey will target over a million objects
chosen from the photometric survey in an attempt to produce a
statistically uniform sample.  The result of the spectroscopic survey
will be a three-dimensional map of the galaxy distribution, in a
volume several orders of magnitude larger than earlier maps.  The
primary targets will be galaxies, selected by a magnitude and surface
brightness limit in the r band. This sample of 900,000 galaxies will
be complemented with 100,000 very red galaxies, selected to include
the brightest galaxies at the cores of clusters. An automated
algorithm will select 100,000 quasar candidates for spectroscopic
follow-up, creating the largest uniform quasar survey to
date. Selected objects from other catalogs will also be targeted.

The spectroscopic observations will be done in overlapping 3$^\circ$
circular tiles. The tile centers are determined by an optimization
algorithm, which maximizes overlaps at areas of highest target
density. The spectroscopic survey will utilize two multi-fiber medium
resolution spectrographs, with a total of 640 optical fibers.  Each
fiber is 3 seconds of arc in diameter, that provide spectral coverage
from 3900 - 9200 \AA. The system can measure 5000 galaxy spectra per
night. The total number of galaxy spectra known to astronomers today
is about 100,000 - only 20 nights of SDSS data!  Whenever the Northern
Galactic cap is not accessible, SDSS repeatedly images several areas
in the Southern Galactic cap to study fainter objects and identify
variable sources.  SDSS has also been developing the software
necessary to process and analyze the data. With construction of both
hardware and software largely finished, the project has now entered a
year of integration and testing. The survey itself will take about 5 
years to complete.

\subsection{The SDSS Archives}

The SDSS will create four main data sets: a photometric catalog, a
spectroscopic catalog, images, and spectra. The photometric catalog is
expected to contain about 500 distinct attributes for each of one
hundred million galaxies, one hundred million stars, and one million
quasars.  These include positions, fluxes, radial profiles, their
errors, and information related to the observations. Each object will
have an associated image cutout ("atlas image") for each of the five
filters. The spectroscopic catalog will contain identified emission
and absorption lines, and one-dimensional spectra for 1 million
galaxies, 100,000 stars, and 100,000 quasars. Derived custom catalogs
may be included, such as a photometric cluster catalog, or quasar
absorption line catalog. In addition there will be a compressed 1TB
Sky Map. These products add up to about 3TB.

The collaboration will release this data to the public after a period
of thorough verification. This public archive is expected to remain
the standard reference catalog for the next several decades.  This
long-lifetime presents design and legacy problems. The design of the
SDSS archival system must allow the archive to grow beyond the actual
completion of the survey. As the reference astronomical data set, each
subsequent astronomical survey will want to cross-identify its objects
with the SDSS catalog, requiring that the archive, or at least a part
of it, be dynamic with a carefully defined schema and metadata.

Observational data from the telescopes is shipped on tapes to Fermi
National Laboratory (FNAL) where it is reduced and stored in the
Operational Archive (OA), protected by a firewall, accessible only to
personnel working on the data processing. Data in the operational
archive is reduced and calibrated via method functions.  Within two
weeks the calibrated data is published to the Science Archive (SA).
The Science Archive contains calibrated data organized for efficient
science use. The SA provides a custom query engine that uses
multidimensional indices. Given the amount of data, most queries will
be I/O limited, thus the SA design is based on a scalable
architecture, ready to use large numbers of cheap commodity servers,
running in parallel.  Science archive data is replicated to Local
Archives (LA) within another two weeks. The data gets into the public
archives (MPA, PA) after approximately 1-2 years of science
verification, and recalibration.  A WWW server will provide public
access.

The Science Archive and public archives employ a three-tiered
architecture: the user interface, an intelligent query engine, and the
data warehouse. This distributed approach provides maximum
flexibility, while maintaining portability, by isolating hardware
specific features. Both the Science Archive and the Operational
Archive are built on top of Objectivity/DB, a commercial OODBMS.

Querying these archives requires a parallel and distributed query
system.  We have implemented a prototype query system.  Each query
received from the User Interface is parsed into a Query Execution Tree
(QET) that is then executed by the Query Engine.  Each node of the QET
is either a query or a set-operation node, and returns a bag of
object-pointers upon execution. The multi-threaded Query Engine
executes in parallel at all the nodes at a given level of the QET.
Results from child nodes are passed up the tree as soon as they are
generated.  In the case of aggregation, sort, intersection and
difference nodes, at least one of the child nodes must be complete
before results can be sent further up the tree.  In addition to
speeding up the query processing, this data push strategy ensures
that even in the case of a query that takes a very long time to
complete, the user starts seeing results almost immediately, or at
least as soon as the first selected object percolates up the tree
(Thakar etal 1999).

\subsection{Typical Queries}

The astronomy community will be the primary SDSS user. They will need
specialized services. At the simplest level these include the
on-demand creation of (color) finding charts, with position
information. These searches can be fairly complex queries on position,
colors, and other parts of the attribute space.  As astronomers learn
more about the detailed properties of the stars and galaxies in the
SDSS archive, we expect they will define more sophisticated
classifications. Interesting objects with unique properties will be
found in one area of the sky. They will want to generalize these
properties, and search the entire sky for similar objects.

A common query will be to distinguish between rare and typical
objects. Other types of queries will be non-local, like "find all the
quasars brighter than r=22, which have a faint blue galaxy within 5
arcsec on the sky". Yet another type of a query is a search for
gravitational lenses: "find objects within 10 arcsec of each other
which have identical colors, but may have a different
brightness". This latter query is a typical high-dimensional query,
since it involves a metric distance not only on the sky, but also in
color space. Special operators are required to perform these queries
efficiently. Preprocessing, like creating regions of attraction is not
practical, given the number of objects, and that the sets of objects
these operators work on are dynamically created by other predicates.

\section{Data Organization}

Given the huge data sets, the traditional Fortran access to flat files
is not a feasible approach for SDSS. Rather non-procedural query
languages, query optimizers, database execution engines, and database
indexing schemes must replace traditional "flat" file processing.
This "database approach" is mandated both by computer efficiency, and
by the desire to give astronomers better analysis tools.

The data organization must support concurrent complex
queries. Moreover, the organization must efficiently use processing,
memory, and bandwidth. It must also support the addition of new data
to the SDSS as a background task that does not disrupt online access.

It would be wonderful if we could use an off-the-shelf SQL, OR, or OO
database system for our tasks, but we are not optimistic that this
will work.  As explained presently, we believe that SDSS requires
novel spatial indices and novel operators. It also requires a dataflow
architecture that executes queries concurrently using multiple disks
and processors. As we understand it, current systems provide few of
these features.  But, it is quite possible that by the end of the
survey, some commercial system will provide these features.  We hope
to work with DBMS vendors towards this end.

\subsection{Spatial Data Structures}

The large-scale astronomy data sets consist primarily of vectors of
numeric data fields, maps, time-series sensor logs and images: the
vast majority of the data is essentially geometric.  The success of
the archive depends on capturing the spatial nature of this
large-scale scientific data.

The SDSS data has high dimensionality -- each item has thousands of
attributes. Categorizing objects involves defining complex domains
(classifications) in this N-dimensional space, corresponding to
decision surfaces.

The SDSS teams are investigating algorithms and data structures to
quickly compute spatial relations, such as finding nearest neighbors,
or other objects satisfying a given criterion within a metric
distance. The answer set cardinality can be so large that intermediate
files simply cannot be created. The only way to analyze such data sets
is to pipeline the answers directly into analysis tools.  This data
flow analysis has worked well for parallel relational database systems
(DeWitt 92).  We expect these data river ideas will link the archive
directly to the analysis and visualization tools.

The typical search of these multi-Terabyte archives evaluates a
complex predicate in k-dimensional space, with the added difficulty
that constraints are not necessarily parallel to the axes. This means
that the traditional indexing techniques, well established with
relational databases, will not work, since one cannot build an index
on all conceivable linear combinations of attributes. On the other
hand, one can use the fact that the data are geometric and every
object is a point in this k-dimensional space (Samet 1990a,b). Data
can be quantized into containers. Each container has objects of
similar properties, e.g. colors, from the same region of the sky. If
the containers are stored as clusters, data locality will be very high
- if an object satisfies a query, it is likely that some of the
object's "friends" will as well. There are non-trivial aspects of how
to subdivide, when the data has large density contrasts (Csabai etal 96).

These containers represent a coarse-grained density map of the
data. They define the base of an index tree that tells us whether
containers are fully inside, outside or bisected by our query. Only
the bisected container category is searched, as the other two are
wholly accepted or rejected. A prediction of the output data volume
and search time can be computed from the intersection.

The SDSS data is too large to fit on one disk or even one server.  The
base-data objects will be spatially partitioned among the servers.  As
new servers are added, the data will repartition. Some of the
high-traffic data will be replicated among servers.  It is up to the
database software to manage this partitioning and replication.  In the
near term, designers will specify the partitioning and index schemes,
but we hope that in the long term, the DBMS will automate this design
task as access patterns change.

There is great interest in a common reference frame the sky that can
be universally used by different astronomical databases. The need for
such a system is indicated by the widespread use of the ancient
constellations - the first spatial index of the celestial sphere. The
existence of such an index, in a more computer friendly form will ease
cross-referencing among catalogs. A common scheme, that provides a
balanced partitioning for all catalogs, may seem to be impossible;
but, there is an elegant solution, a 'shoe that fits all': that
subdivides the sky in a hierarchical fashion. Our approach is
described in detail by Kunszt etal (1999).

\subsection{Broader Metadata Issues}

There are several issues related to metadata for astronomy
datasets. One is the database schema within the data warehouse,
another is the description of the data extracted from the archive and
the third is a standard representation to allow queries and data to be
interchanged among several archives. The SDSS project uses Platinum
Technology's Paradigm Plus, a commercially available UML tool, to
develop and maintain the database schema. The schema is defined in a
high level format, and a script generator creates the .h
files for the C++ classes, and the .ddl files for Objectivity/DB. This
approach enables us to easily create new data model representations in
the future (SQL, IDL, XML, etc).

About 20 years ago, astronomers agreed on exchanging most of their
data in self-descriptive data format. This format, FITS, standing for
the Flexible Image Transport System (Wells 81) was primarily designed
to handle images. Over the years, various extensions supported more
complex data types, both in ASCII and binary form. FITS format is well
supported by all astronomical software systems. The SDSS pipelines
exchange most of their data as binary FITS files. Unfortunately, FITS
files do not support streaming data, although data could be blocked
into separate FITS packets. We are currently implementing both an
ASCII and a binary FITS output stream, using such a blocked
approach. We expect large archives to communicate with one another via
a standard, easily parseable interchange format. We plan to define the
interchange formats in XML, XSL, and XQL.

The Operational Archive exports calibrated data to the Science Archive
as soon as possible. Datasets are sent in coherent chunks. A chunk
consists of several segments of the sky that were scanned in a single
night, with all the fields and all objects detected in the
fields. Loading data into the Science Archive could take a long time
if the data were not clustered properly. Efficiency is important,
since about 20 GB will be arriving daily. The incoming data are
organized by how the observations were taken. In the Science Archive
they will be inserted into the hierarchy of containers as defined by
the multi-dimensional spatial index, according to their colors and
positions.

Data loading might bottleneck on creating the clustering
units - databases and containers - that hold the objects. Our load design
minimizes disk accesses, touching each clustering unit at most once
during a load.  The chunk data is first examined to construct an
index. This determines where each object will be located and creates a
list of databases and containers that are needed. Then data is
inserted into the containers in a single pass over the data objects.

\subsection{Scalable Server Architectures}

Accessing large data sets is primarily I/O limited. Even with the best
indexing schemes, some queries must scan the entire data
set. Acceptable I/O performance can be achieved with expensive,
ultra-fast storage systems, or with many of commodity servers
operating in parallel. We are exploring the use of commodity servers
and storage to allow inexpensive interactive data analysis.  We are
still exploring what constitutes a balanced system design: the
appropriate ratio between processor, memory, network bandwidth, and
disk bandwidth.

Using the multi-dimensional indexing techniques described in the
previous section, many queries will be able to select exactly the data
they need after doing an index lookup.  Such simple queries will just
pipeline the data and images off of disk as quickly as the network can
transport it to the astronomer's system for analysis or visualization.
When the queries are more complex, it will be necessary to scan the
entire dataset or to repartition it for categorization, clustering,
and cross comparisons.  Experience will teach us the ratio
between processor power, memory size, IO bandwidth, and
system-area-network bandwidth.

Our simplest approach is to run a scan machine that continuously scans
the dataset evaluating user-supplied predicates on each object
(Acharya 95).  Consider building an array of 20 nodes, each with 4
Intel Xeon 450 Mhz processors, 256MB of RAM, and 12x18GB disks (4TB of
storage in all). Experiments show that one such node is capable of
reading data at 150 MBps while using almost no processor time (Hartman
99).  If the data is spread among the 20 nodes, they can scan the data
at an aggregate rate of 3 GBps. This half-million dollar system could
scan the complete (year 2004) SDSS catalog every 2 minutes.  By then
these machines should be 10x faster.  This should give
near-interactive response to most complex queries that involve
single-object predicates.

Many queries involve comparing, classifying or clustering objects. We
expect to provide a second class of machine, called a hash machine
that performs comparisons within data clusters.  Hash machines
redistribute a subset of the data among all the nodes of the
cluster. Then each node processes each hash bucket at that node.  This
parallel-clustering approach has worked extremely well for relational
databases in joining and aggregating data.  We believe it will work
equally well for scientific spatial data.

The hash phase scans the entire dataset, selects a subset of the
objects based on some predicate, and "hashes" each object to the
appropriate buckets - a single object may go to several buckets (to
allow objects near the edges of a region to go to all the neighboring
regions as well).  In a second phase all the objects in a bucket are
compared to one another.  The output is a stream of objects with
corresponding attributes.

These operations are analogous to relational hash-join, hence the name
(DeWitt 92).  Like hash joins, the hash machine can be highly parallel,
processing the entire database in a few minutes. The application of
the hash-machine to tasks like finding gravitational lenses or
clustering by spectral type or by redshift-distance vector should be
obvious: each bucket represents a neighborhood in these
high-dimensional spaces. We envision a non-procedural programming
interface to define the bucket partition and analysis functions.

The hash machine is a simple form of the more general data-flow
programming model in which data flows from storage through various
processing steps.  Each step is amenable to partition parallelism. The
underlying system manages the creation and processing of the flows.
This programming style has evolved both in the database community
(DeWitt 92, Graefe 93, Barclay 95) and in the scientific programming
community with PVM and MPI (Gropp 98).  This has evolved to a general
programming model as typified by a river system (Arpaci-Dusseau 99).

We propose to let astronomers construct dataflow graphs where the
nodes consume one or more data streams, filter and combine the data,
and then produce one or more result streams.  The outputs of these
rivers either go back to the database or to visualization
programs. These dataflow graphs will be executed on a river-machine
similar to the scan and hash machine.  The simplest river systems are
sorting networks.  Current systems have demonstrated that they can
sort at about 100 MBps using commodity hardware and 5 GBps if using
thousands of nodes and disks (Sort benchmark).

With time, each astronomy department will be able to afford local
copies of these machines and the databases, but for now, they will be
a network service. The scan machine will be interactively scheduled:
when an astronomer has a query, it is added to the query mix
immediately.  All data that qualifies is sent back to the astronomer,
and the query completes within the scan time.  The hash and river
machines will be batch scheduled.

\subsection{Desktop Data Analysis}

Most astronomers will not be interested in all of the hundreds of
attributes of each object.  Indeed, most will be interested in only
10\% of the entire dataset - but different communities and individuals
will be interested in a different 10\%. We plan to isolate the 10 most
popular attributes (3 Cartesian positions on the sky, 5 colors, 1
size, 1 classification parameter) into small 'tag' objects, which
point to the rest of the attributes. Then we will build a spatial
index on these attributes.  These will occupy much less space, thus
can be searched more than 10 times faster, if no other attributes are
involved in the query.

Large disks are available today, and within a few years 100GB disks
will be common. This means that all astronomers can have a vertical
partition of the 10\% of the SDSS on their desktops.  This will be
convenient for targeted searches and for developing algorithms.  But,
full searchers will still be much faster on the server machines
because the servers will have much more IO bandwidth and processing
power. Vertical partitioning can also be applied by the scan, hash,
and river machines to reduce data movement and to allow faster scans
of popular subsets. We also plan to offer a 1\% sample (about 10 GB)
of the whole database that can be used to quickly test and debug
programs.  Combining partitioning and sampling converts a 2 TB data
set into 2 gigabytes, which can fit comfortably on desktop
workstations for program development.

It is obvious, that with multi-terabyte databases, not even the
intermediate data sets can be stored locally. The only way this data
can be analyzed is for the analysis software to directly communicate
with the Data Warehouse, implemented on a server cluster, as discussed
above. Such an Analysis Engine can then process the bulk of the raw
data extracted from the archive, and the user needs only to receive a
drastically reduced result set.

Given all these efforts to make the server parallel and distributed,
it would be inefficient to ignore IO or network bottlenecks at the
analysis level. Thus it is obvious that we need to think of the
analysis engine as part of the distributed, scalable computing
environment, closely integrated with the database server itself. Even
the division of functions between the server and the analysis engine
will become fuzzy - the analysis is just part of the river-flow
described earlier. The pool of available CPU's will be allocated to
each task.

The analysis software itself must be able to run in parallel. Since it
is expected that scientists with relatively little experience in
distributed and parallel programming will work in this environment, we
need to create a carefully crafted application development
environment, to aid the construction of customized analysis
engines. Data extraction needs to be considered also carefully. If our
server is distributed and the analysis is on a distributed system, the
extracted data should also go directly from one of the servers to one
of the many Analysis Engines. Such an approach will also distribute
the network load better.

\section{Summary}

Astronomy is about to be revolutionized by having a detailed atlas of
the sky available to all astronomers.  With the SDSS archive it will
be easy for astronomers to pose complex queries to the catalog and get
answers within seconds, and within minutes if the query requires a
complete search of the database.  The SDSS datasets pose interesting
challenges for automatically placing and managing the data, for
executing complex queries against a high-dimensional data space, and
for supporting complex user-defined distance and classification
metrics. The SDSS project is "riding Moore's law": the data set we
started to collect today - at a linear rate - will be much more
manageable tomorrow, with the exponential growth of CPU speed and
storage capacity. The scalable archive design presented here will be
able to adapt to such changes.

\acknowledgments

We would like to acknowledge support from the Astrophysical Research
Consortium, the HSF, NASA and Intel's Technology for Education 2000
program, in particular George Bourianoff (Intel).


\end{document}